\documentclass{jjap3}
\usepackage{bm}
\usepackage[dvipdfm]{graphicx}

\newcommand{\rbk}[1]{\left( #1 \right)}

\newcommand{\sbk}[1]{\left[ #1 \right]}
\newcommand{\abk}[1]{\left\langle #1 \right\rangle}

\newcommand{\retn}{\nonumber \\ }

\newcommand{\pdif}[2]{\frac{\partial #1}{\partial #2}}

\newcommand{\percent}{\%~}

\title{ Molecular Dynamics Simulation of Chemical Vapor Deposition of Amorphous Carbon: Dependence on H/C Ratio of Source Gas}

\author{Atsushi M. {Ito}$^{1}$\thanks{E-mail address: ito.atsushi@nifs.ac.jp},
 Arimichi {Takayama}$^{1}$, Seiki {Saito}$^{2}$, Noriyasu {Ohno}$^{3}$, Shin {Kajita}$^3$, and Hiroaki {Nakamura}$^{1,2}$}

\inst{$^{1}$Department of Helical Plasma Research, National Institute for Fusion Science, 322-6 Oroshicho, Toki, Gifu 509-5292, Japan\\
$^{2}$Department of Energy Engineering and Science, Nagoya University, Furocho, Chikusa-ku, Nagoya 464-8602, Japan\\
$^{3}$Ecotopia Science Institute, Nagoya University, Furocho, Chikusa-ku, Nagoya 464-8602, Japan}

\abst{%
By molecular dynamics simulation, the chemical vapor deposition of amorphous carbon onto graphite and diamond surfaces was studied. In particular, we investigated the effect of source H/C ratio, which is the ratio of the number of hydrogen atoms to the number of carbon atoms in a source gas, on the deposition process. In the present simulation, the following two source gas conditions were tested: one was that the source gas was injected as isolated carbon and hydrogen atoms, and the other was that the source gas was injected as hydrocarbon molecules. Under the former condition, we found that as the source H/C ratio increases, the deposition rate of carbon atoms decreases exponentially. This exponential decrease in the deposition rate with increasing source H/C ratio agrees with experimental data. However, under the latter molecular source condition, the deposition rate did not decrease exponentially because of a chemical reaction peculiar to the type of hydrocarbon in the source gas.}

\begin{document}
\maketitle

\section{Introduction} 

Carbon is the most widely used material structure among all elements. It has two stable lattice structures, graphite and diamond. On the nanoscale, graphene, carbon nanotubes, carbon nanowalls, fullerene, and carbon onion are considered examples of nanostructures. The reason why carbon has many kinds of nanostructures is that sp$^2$ and sp$^3$ bonding states are balanced in binding energy. Silicon cannot form these kinds of nanostructure because its sp hybrid orbital consists of 3s and 3p orbitals, and then the sp$^2$ bonding state is weaker than the sp$^3$ bonding state.

Amorphous structures are difficult to classify since exact quantities that can be used to characterize them have not yet been found. In particular, amorphous carbon is a complex material because of the sp$^2$ and sp$^3$ bonding states. In addition, amorphous carbon mixed with hydrogen atoms is a more complex material. Thus, amorphous carbon has been investigated in numerous works \cite{Jariwala, Peter}. Chemical vapor deposition (CVD) and plasma CVD are often used in experiments. Diamond-like carbon (DLC) is considered as a useful material; thus, its formation process is actively investigated \cite{Robertson0, Nakao, Merel, Mokuno} in engineering. However, the formation processes of the amorphous carbon and DLC by CVD are not known well.

In the study of nuclear fusion, the amorphous carbon obtained by deposition, which is similar to the plasma CVD process, is disadvantageous. In nuclear fusion devices, hydrogen plasma causes chemical erosion and chemical sputtering on divertor plates composed of carbon materials. Generated hydrocarbon molecules are transported by edge plasma, whose deposition creates amorphous carbon with hydrogen atoms \cite{Skinner}. The above phenomenon between plasma and surfaces is called plasma surface interaction (PSI). The PSI of the combination of plasma and carbon materials appears in various applications, such as the processing, coating, and growth of thin films. We have investigated PSI on carbon materials with hydrogen atoms by molecular dynamics (MD) simulation with the clarification of nanoscale dynamics. In our previous works, the destruction mechanisms of the surface in PSI, especially chemical sputtering and chemical erosion, were investigated \cite{Ito_IAEA, Nakamura_PET}. In this study, we started to investigate the deposition process of a hydrocarbon gas onto carbon materials. On the nanoscale, the deposition in nuclear fusion devices and the CVD in laboratory experiments are considered to have similar mechanisms.

Our simulation method and model are described in \S 2.
In \S 3, we present and discuss the simulation results.
This paper ends with conclusions in \S 4.

\section{Simulation Model}

\subsection{Molecular dynamics for plasma surface interaction}

MD simulation for PSI is generally performed as follows. Figure \ref{fig:method1} shows the conceptual diagram. A rectangular simulation box is considered. All particles move in the simulation box only. A target material is placed at the bottom of the simulation box. To make the surface of the target material vertical to the $z$-axis, the simulation box is placed relative to the coordination axes. The top and bottom of the simulation box are defined as positive and negative sides in the $z$ coordinate. Incident particles are injected from the top side into the surface of the target material, where the $z$ components of momentum of the injected particles are negative. The sizes of the simulation box in the $x$- and $y$-directions cannot be set arbitrarily because periodic boundary conditions in the $x$- and $y$-directions are imposed on the simulation box. Therefore, the sizes of the simulation box in the $x$- and $y$-directions are determined as the lengths of the unit cells of integer pieces. If the surface slants toward the lattice direction, the unit cells are located in the simulation box at the slant. When the target material is selected, the species of atom composition, the lattice structure, and the Miller indices are determined. If the Miller indices are large, the size of the simulation box increases to line up the unit cells of integer pieces. The bottom of the target material should be fixed to the simulation box because the target material is entirely pushed by the incident particles. When the target material is a carbon material, flat (0001), armchair (11$\bar{2}$0), and zigzag (10$\bar{1}$0) surfaces of graphite, and diamond (100), (110), and (111) surfaces are typically used. The target material with an amorphous structure can be used after an amorphous material is created by the presimulation of the deposition or melting of a crystalline material. 

How physical phenomena can be treated strongly depends on the conditions of incident particles. If the incident energy is high, incident particles penetrate the target material deeply, which brings about the retention of particles, or they beat out particles in the surface area, which is called physical sputtering. Even if the incident energy is low, the covalent bonds in the surface area are broken by a chemical reaction with incident particles, which is called chemical sputtering or chemical erosion. It is difficult to classify the chemical sputtering and chemical erosion because the chemical sputtering does not beat out any atoms. If an incident particle includes an atom that can have many covalent bonds, e.g., a carbon atom, it creates deposits on the surface.

\subsection{Algorithm}

To deal with chemical reactions in MD, the modified Brenner reactive empirical bond order (REBO) potential was employed \cite{Brenner2002, Ito_jpsj, Ito_ICNSP20}.
The modified Brenner REBO potential is given by
\begin{eqnarray}
	U &=& \sum_{i,j>i} \sbk{V_{[ij]}^\mathrm{R}( r_{ij} )
		 - \bar{b}_{ij}(\{r\},\{\theta^\mathrm{B}\},\{\theta^\mathrm{DH}\}) V_{[ij]}^\mathrm{A}(r_{ij}) },
		\label{eq:model_rebo} \retn
		V_{[ij]}^\mathrm{R}( r_{ij}) &=& f_{[ij]}^\mathrm{c}(r_{ij}) \rbk{1 + \frac{Q_{[ij]}}{r_{ij}}}
		 A_{[ij]} \exp\rbk{-\alpha_{[ij]} r_{ij}}, \\
	V_{[ij]}^\mathrm{A}( r_{ij}) &=& f_{[ij]}^\mathrm{c}(r_{ij}) \sum_{n = 1}^3 B_{n[ij]} \exp\rbk{-\beta_{n[ij]} r_{ij}},
\end{eqnarray}
where $r_{ij}$ is the distance between the $i$-th and $j$-th atoms.
The functions $V_{[ij]}^{\mathrm{R}}$ and $V_{[ij]}^{\mathrm{A}}$ correspond to repulsion and attraction, respectively.
The function $\bar{b}_{ij}$ generates a multibody force.
The modified Brenner REBO potential can treat structures peculiar to carbon, i.e., $\textrm{sp}^2$ and $\textrm{sp}^3$ bonds.
The simulation time is developed using the second-order symplectic integration\cite{Suzuki};
the time step is $5.09 \times 10^{-18} \textrm{~s}$.
This time step is smaller than that of a general MD simulation.
The modified Brenner REBO potential is represented by the interpolation between the potential of the bonding state and the potential of the unbonding state using cutoff functions, which have two cutoff lengths.
Because the interpolation interval ranges from 0.3 to 0.6~\AA, a small time step is required.

In this simulation, the material temperature was controlled using the Langevin equation:
\begin{equation}
	\dot{p}_i = -\pdif{U}{r_i} - \gamma p_i + \xi_i(t),
	\label{eq:1}
\end{equation}
where $r_i$ and $p_i$ are the position and momentum of a carbon atom, respectively.
The coefficient of friction $\gamma \geq 0$.
The random thermal force $\xi_i(t)$ satisfies 
\begin{equation}
	\abk{\xi_i(t)} = 0, \\
	\abk{\xi_i(t)\xi_j(t')} = 2D\delta(t-t')\delta_{ij},
	\label{eq:2}
\end{equation}
where the symbol $\abk{\cdots}$ indicates the expected value of a random variable, $\delta(t-t')$ is the Dirac delta function, and $\delta_{ij}$ is the Kronecker delta.
The magnification of the random thermal force $D$ is given by the Einstein relation
\begin{equation}
	D = \gamma m_i k_\mathrm{B} T_\mathrm{s},
	\label{eq:3}
\end{equation}
where $m_i = 12$ u is the mass of a carbon atom, $k_\mathrm{B}$ is the Boltzmann constant, and $T_\mathrm{s}$ is the setting temperature of a material.
In the numerical simulation with the time step $\Delta t$, we substitute the difference equation
\begin{equation}
	p_i\rbk{ t + \frac{\Delta t}{2}} = p_i\rbk{ t - \frac{\Delta t}{2}} - \sbk{\pdif{U}{r_i}(t) - \gamma p_i\rbk{ t - \frac{\Delta t}{2}} }\Delta t + \sqrt{2D\Delta t}B_i
	\label{eq:4}
\end{equation}
for eq. (\ref{eq:1}).
The normally distributed random number $B_i$ is generated by the Box-Muller transform \cite{Box-Muller} from the uniformly distributed random number generated by single instruction multiple data (SIMD) oriented Fast Mersenne Twister \cite{Saito2006}. In the present simulations, the thermal relaxation time $1/\gamma$ was 0.1 ps.

To investigate CVD by MD simulation, the simulation system was prepared as follows. The target materials were graphite and diamond. The surface of the graphite material was a flat (0001) surface of 2.00 x 2.17 nm$^2$. The graphite material had a layer structure composed of four graphene sheets, which were stacked to form the "ABAB" pattern. Each graphene sheet consisted of 160 carbon atoms. The interlayer distance of graphite materials was set to 3.35 \AA~initially. The bottommost graphene sheet was regarded as the fixed layer, and then the 160 carbon atoms in the fixed layer were bound by harmonic potentials whose centers were the initial positions of the atoms. Diamond materials with the (100) and (111) surfaces of 2.14 x 2.14 and 2.02 x 2.19 nm$^2$, respectively, were used. These diamond materials consisted of 648 and 640 carbon atoms, respectively. This difference in the number of carbon atoms made the sizes of the surfaces similar. In the fixed layers of these diamond materials,  72 and 80 carbon atoms were respectively bound by harmonic potentials.
The graphite and diamond materials were connected to the Langevin thermostat. The kinetic energies of the atoms, which were the component atoms of the initial materials, were controlled by the thermostat, while those of the atoms injected into the surfaces were not controlled by the thermostat even if they were deposited on the surfaces.

As the precursors of deposition, carbon atoms, hydrogen atoms, and hydrocarbon molecules were injected into the surfaces. The flux of the carbon atoms was 2.5 x 10$^{30}$ atoms/m$^2$s. The hydrogen atoms were always injected with the carbon atoms and the fluxes of the hydrogen atoms were from 2.5 x 10$^{29}$ to 2.0 x 10$^{31}$ atoms/m$^2$s. The fluxes of hydrocarbon molecules were set to have the same value in terms of the number of carbon atoms. For example, the fluxes of CH$_2$ and C$_2$H$_2$ were 2.5 x 10$^{30}$ and 1.25 x 10$^{30}$ atoms/m$^2$s, respectively. The incident energy was 1.0 eV.

\section{Results and Discussion}

First, only carbon atoms were injected into the diamond and graphite surfaces to compare the deposition mechanisms.
All carbon deposits on the diamond and graphite surfaces grew with elapsed time linearly. The growth rates of the carbon deposits on the graphite (0001) surface and the diamond (100) and (111) surfaces were almost the same. The carbon deposits consisted of about 80\percent of sp$^2$ carbon atoms, 16\percent of sp$^3$ carbon atoms, and 5-8\percent of sp$^1$ carbon atoms.
Although the sp$^3$ carbon atoms were fewer than sp$^2$ carbon atoms, almost all the sp$^3$ carbon atoms were connected to each other (Fig. \ref{fig:cvd6}).
The details of the compositions of the carbon deposits are shown in Table \ref{t1}. From the composition ratio, the carbon deposits produced by MD simulation were regarded as amorphous carbon (a-C). The carbon deposits were probably called graphite-like amorphous carbon, which is mainly composed of sp$^2$ carbon atoms. They were not diamond-like carbon (DLC), which has sp$^3$ carbon atoms of more than 80\percent\cite{Robertson1, Robertson2, McKenzie, Fallon, Pharr}. The density of the carbon deposits was about 2.4 g/cm$^3$, which is higher than that of graphite but lower than that of DLC. These composition ratios and densities agreed with those of the amorphous carbon produced by tight binding MD \cite{Wang}.

To examine the dependence of the deposition mechanisms on the source H/C ratio, we performed MD simulation in which carbon and hydrogen were simultaneously injected as isolated atoms into the surfaces at a material temperature of 600 K. The ratio of incident flux of hydrogen atoms to that of carbon atoms was defined as the source H/C ratio. The incident flux of carbon atoms was always fixed to 2.5 x 10$^{30}$ atoms/m$^2$s, and the incident flux of hydrogen atoms was selected to be 0.1 to 8.0 times that of the carbon atoms. For example, when the flux of the hydrogen atoms was 1.25 x 10$^{30}$ atom/m$^2$s, H/C = 0.5. In this paper, the deposition rate was defined as the ratio of the number of deposited carbon atoms to the number of injected carbon atoms. Figure \ref{fig:cvd2kai} shows the deposition rate as a function of the source H/C ratio. The deposition rates plotted in this figure were calculated at the moment when 2500 carbon atoms were injected. Because the surface area was about 4 nm$^2$, the number of carbon atoms corresponding to a pure monolayer was 160. Then, a deposition rate less than 0.064 $ ( = 160 / 2500 )$ indicates that the deposited carbon atoms could not form a sufficient amount of deposition layer, although the density of amorphous carbon is smaller than that of the pure crystal, strictly speaking. From this figure, we propose that the deposition rate of carbon atom $D_\mathrm{C}$ can be fitted by the exponential function of the source H/C ratio $r_\mathrm{H/C}$ as follows:
\begin{eqnarray}
	D_\mathrm{C} =  D_0 \exp\rbk{- \frac{r_\mathrm{H/C}}{r_0}},
\label{eq:cvd1}
\end{eqnarray}
where the parameters (D$_0$, r$_0$) were (0.698, 0.642), (0.663,0.6370), and (0.665, 0.441) for the diamond (100) and (111), and graphite (0001) surfaces, respectively. The parameter R$_0$ was the deposition rate when only the carbon atom was the source atom. The deposition rates on the diamond (100) and (111) surfaces were almost the same. The deposition rate on the graphite (0001) surface was smaller than those of the diamond surfaces.
In plasma CVD experiments \cite{Liu, Schwarz}, the relationship between deposition rate and source H/C ratio was researched. Actually, we confirmed that the deposition rate in the experimental data cited from Liu et al.[Fig. \ref{fig:cvd2kai}(c)]\cite{Liu}, which was defined as the deposition thickness per unit time, can be fitted by the exponential function of source H/C ratio.

The compositions of the carbon deposits for the bonding state of carbon atoms were calculated as functions of the source H/C ratio, which are shown in Fig. \ref{fig:cvd3}. As the source H/C ratio increased, the number of sp$^2$ carbon atoms decreased but those of sp$^1$ and sp$^3$ carbon atoms increased slightly. In general, it is considered that sp$^2$ carbon atoms catch hydrogen atoms and change to sp$^3$ carbon atoms in a hydrogen-mixed layer. From the result, it is clear that hydrogen termination formed sp$^1$ carbon atoms. That is, sp$^1$ carbon atoms that have one C-H bond and one C-C bond were generated as carbon chain molecules, whose ends of carbon atoms are terminated by hydrogen atoms. The formation process from carbon and hydrogen atoms may prefer the use of the sp$^1$ carbon atom between C-C and C-H bonds. Actually, in hydrocarbon formation in the gas phase, carbon chain molecules are formed first. However, after carbon chain molecules are formed in the gas phase, they change into ring and cluster structures slowly. From this fact about the formation process in the gas phase, if the present deposition simulation is executed for a long time at a lower incident flux, sp$^1$ carbon atoms can potentially change into sp$^2$ or sp$^3$ carbon atoms.

In the deposition of the amorphous carbon material, the source gas not fully resolved into atoms, the deposited particle could have possibly been molecules or radicals. Then, the deposition using C$_2$, CH, CH$_2$, C$_2$H$_2$, and carbon chain C$_4$H$_2$ molecules was tested. Figure \ref{fig:cvd4} shows {the deposition rates of carbon atoms in the cases of these hydrocarbon molecules.} The deposition rate of C$_2$ was greater than that of the carbon atom, while the deposition rates of the hydrocarbon molecules were smaller. C$_2$H$_2$ hardly formed carbon deposits on all the surfaces. For CH and CH$_2$, the deposition rates on the diamond (100) surface were greater than that on the diamond (111) surface. Only on the graphite (0001) surface, the deposition amount of CH$_2$ was passably larger than that of CH. Here, the {reason why} CH$_2$ could deposit more easily than CH although the source H/C ratio of CH$_2$ was greater than that of CH was determined by comparing CH$_2$ deposition with CH deposition. {Figure \ref{fig:cvd5} shows the deposition process actually observed in the present MD simulation. The photograph of CH injection in Fig. \ref{fig:cvd5}(a) illustrates many C$_2$H residues, whose two carbon atoms are in the sp$_1$ state, on the graphite surface. A typical CH deposition process is as follows [Fig. \ref{fig:cvd5}(c)]. First, a CH molecule was adsorbed on the surface and formed a CH residue. Next, the second CH molecule reacted with the CH residue and then a C$_2$H residue was generated. The C$_2$H residue could not grow because the hydrogen atom on top of the C$_2$H residue blocked the third CH molecule. On the other hand, the photograph of CH$_2$ injection in Fig. \ref{fig:cvd5}(b) shows many sp$^3$ carbon atoms that have two C-C bonds and two C-H bonds. In the CH$_2$ deposition process [Fig. \ref{fig:cvd5}(d)], CH$_2$ molecules form a polymer-like structure on the graphite surface. Because the polymer-like structure had alternating \textit{trans} and \textit{cis} isomers, the structure was not straight. That is, the hydrogen atom, which reflected incoming source particles, was not located on only the top of the polymer. Therefore, CH$_2$ could be adsorbed on top of the polymer one after another.}

\begin{table}[tb]
\caption{Compositions of deposits.}
\label{t1}
\begin{tabular}{ccccc}
\hline
Surface & sp$^1$ (\%) & sp$^2$ (\%) & sp$^3$ (\%) & Density (g/cm$^3$)\\
\hline
Diamond (100)   & 6.5 & 77.3 & 16.2 & 2.49 \\
Diamond (111)   & 7.5 & 75.8 & 16.7 & 2.34\\
Graphite (0001) & 4.8 & 78.7 & 16.5 & 2.42\\
\hline
\end{tabular}
\end{table}

\section{Conclusions}

The chemical vapor deposition of amorphous carbon was investigated by MD simulation. The base surfaces of the present simulation were the graphite (0001) and diamond (100) and (111) surfaces, and the source gas was injected in the following two states: isolated carbon and hydrogen atoms being injected according to the source H/C ratios of 0.0 to 8.0, and hydrocarbon molecules being injected as  CH, CH$_2$, C$_2$H$_2$, C$_4$H$_2$, and C$_2$. We determined the dependence of the source H/C ratio on the deposition process of amorphous carbon. As a result, in the isolated atom injection, we found that the deposition rate of carbon atoms can be fitted by the exponential function of the source H/C ratio. It is confirmed that the experimental data cited from Liu et al.\cite{Liu} also agree with this exponential decrease in the deposition rate with increasing source H/C ratio. However, in the hydrocarbon molecular injection, the deposition rate exponentially decreases as the source H/C ratio increases. As an example of the breaking of the exponential decrease, we chose the MD simulation result showing that CH$_2$ was more rapidly deposited than CH although the H/C ratio of CH$_2$ was greater than that of CH. We expect that the exponential decrease in the deposition rate with increasing source H/C ratio tends toward macroscopic modeling for CVD of amorphous carbon.

\section*{Acknowledgments}
Numerical simulations were carried out by the use of the Plasma Simulator at the National Institute for Fusion Science. The work is supported by the National Institutes of Natural Sciences undertaking Forming Bases for Interdisciplinary and International Research through Cooperation Across Fields of Study and Collaborative Research Program (No. NIFS09KEIN0091), a Grant-in-Aid for JSPS Fellows (No. 20-3829) and a Grant-in-Aid for Scientific Research (No. 19055005) from the Ministry of Education, Culture, Sports, Science and Technology.


\newpage

\noindent \textbf{Figure Captions}\\

\begin{figure}
	\centering
	\resizebox{\linewidth}{!}{\includegraphics{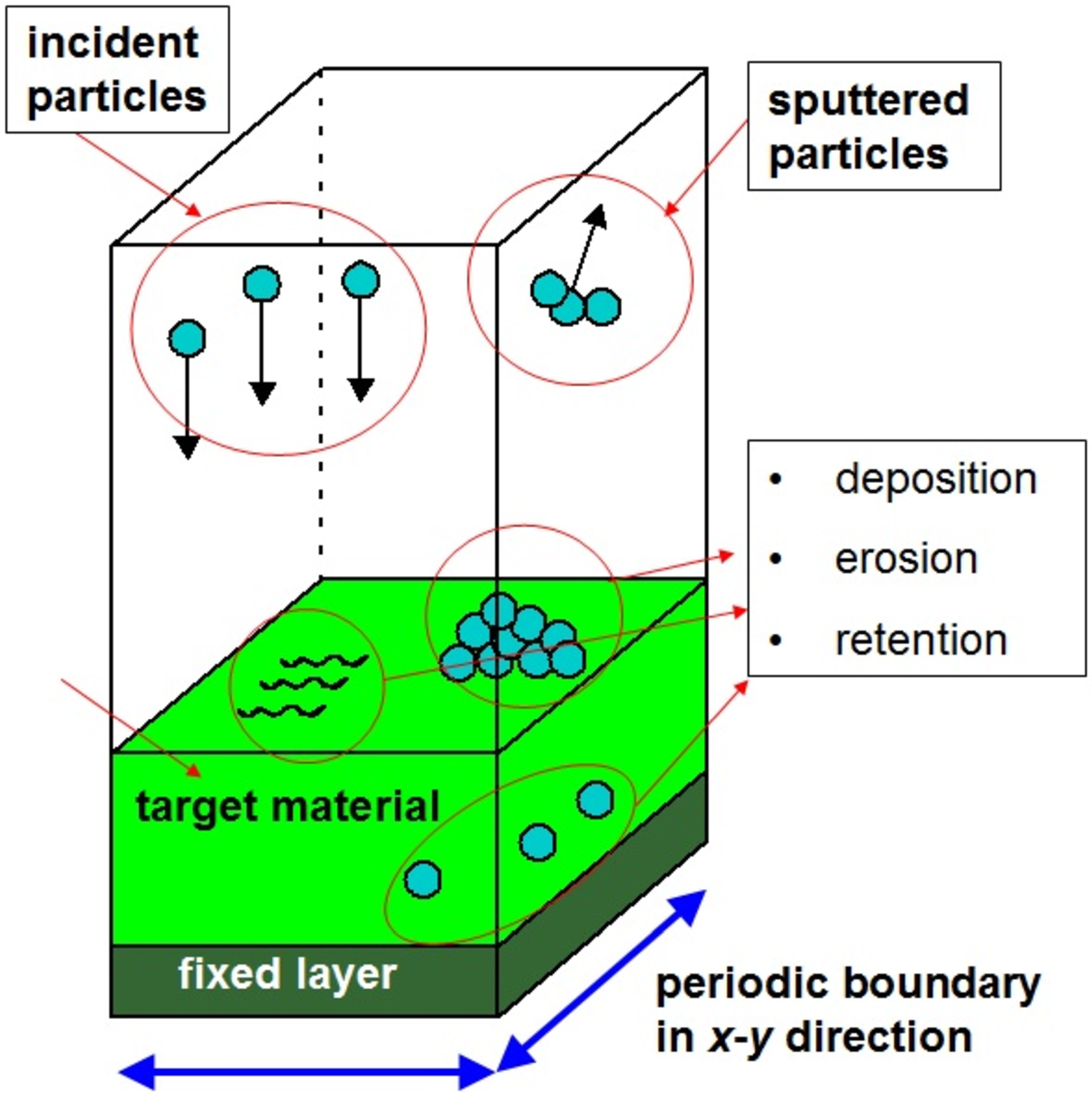}}
	\caption{~}
	\label{fig:method1}
\end{figure}

\begin{figure}
	\centering
		\resizebox{!}{\linewidth}{\includegraphics{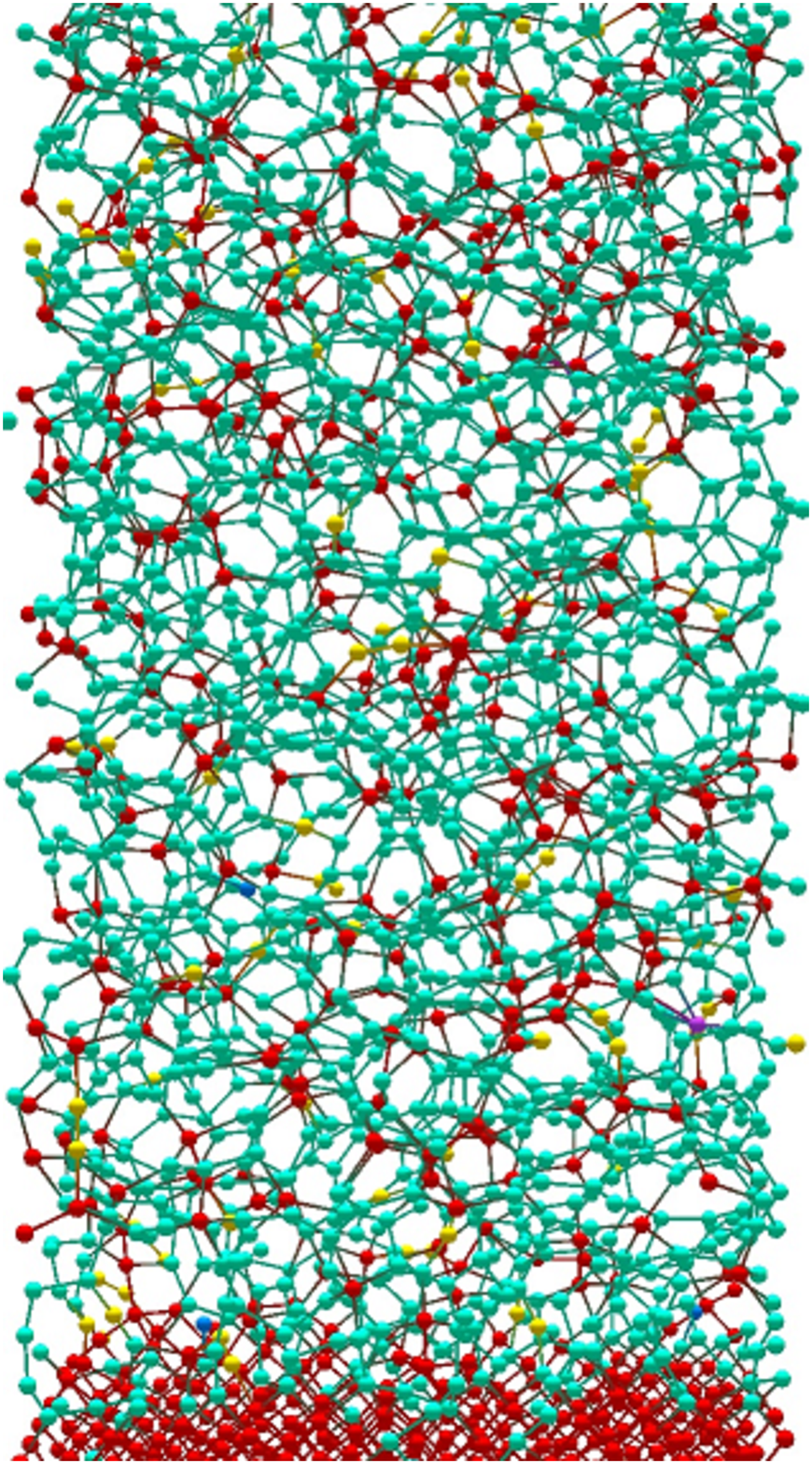}}
	\caption{~}
	\label{fig:cvd6}
\end{figure}

\begin{figure*}
\begin{tabular}{ccc}
		
		\resizebox{0.3\linewidth}{!}{\includegraphics{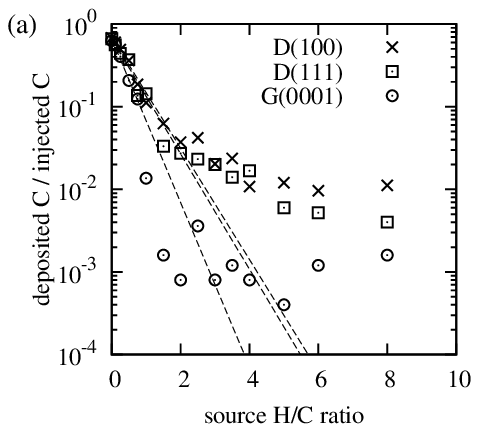}} &
		\resizebox{0.3\linewidth}{!}{\includegraphics{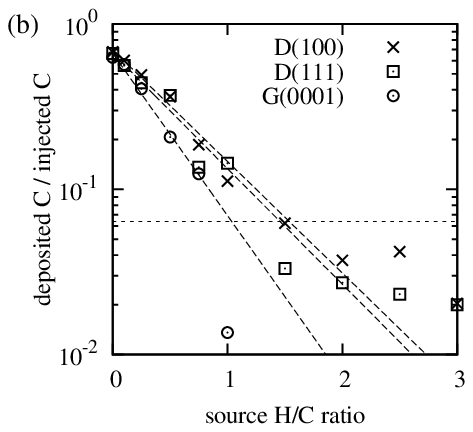}} &
		\resizebox{0.3\linewidth}{!}{\includegraphics{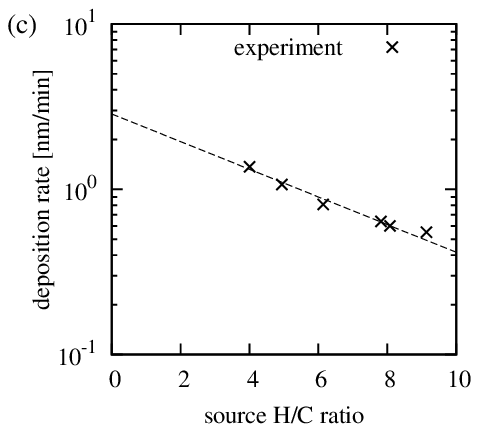}} \\
		
	\end{tabular}
	\caption{~}
\label{fig:cvd2kai}
\end{figure*}

\begin{figure*}
\begin{tabular}{ccc}
		
		\resizebox{0.3\linewidth}{!}{\includegraphics{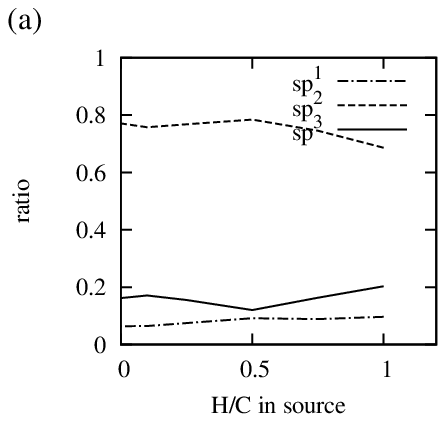}} &
		\resizebox{0.3\linewidth}{!}{\includegraphics{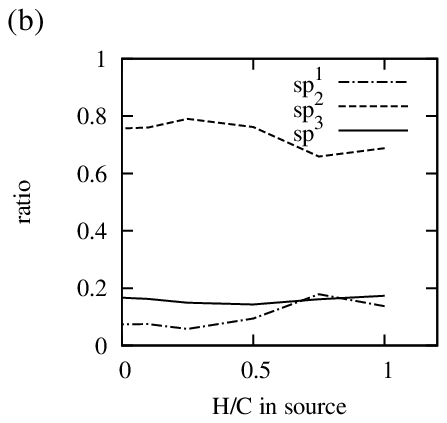}} &
		\resizebox{0.3\linewidth}{!}{\includegraphics{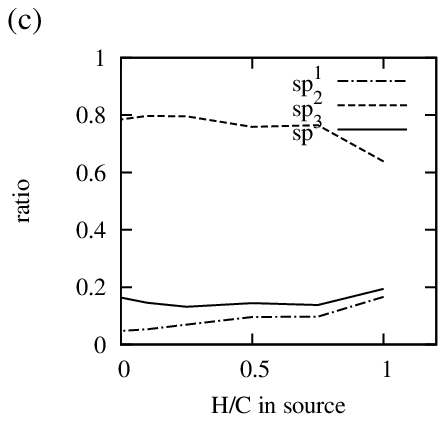}} \\
		
	\end{tabular}
	\caption{~}
\label{fig:cvd3}
\end{figure*}

\begin{figure*}
\begin{tabular}{ccc}
		
		\resizebox{0.3\linewidth}{!}{\includegraphics{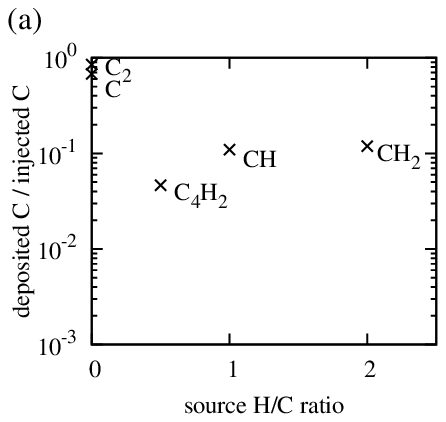}} &
		\resizebox{0.3\linewidth}{!}{\includegraphics{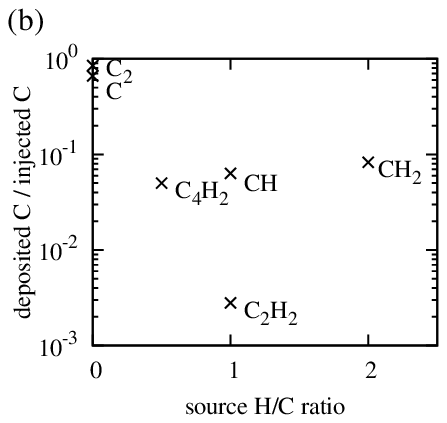}} &
		\resizebox{0.3\linewidth}{!}{\includegraphics{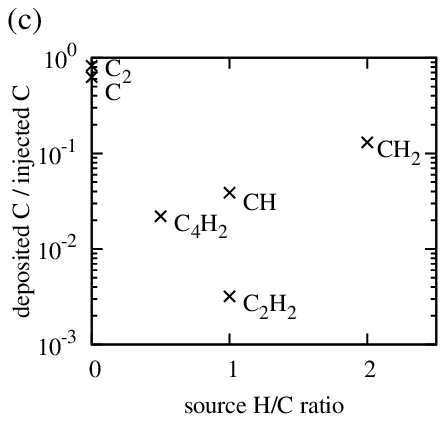}} \\
		
	\end{tabular}
	\caption{~}
\label{fig:cvd4}
\end{figure*}

\begin{figure}
	\centering
\begin{tabular}{cc}
		\resizebox{0.5\linewidth}{!}{\includegraphics{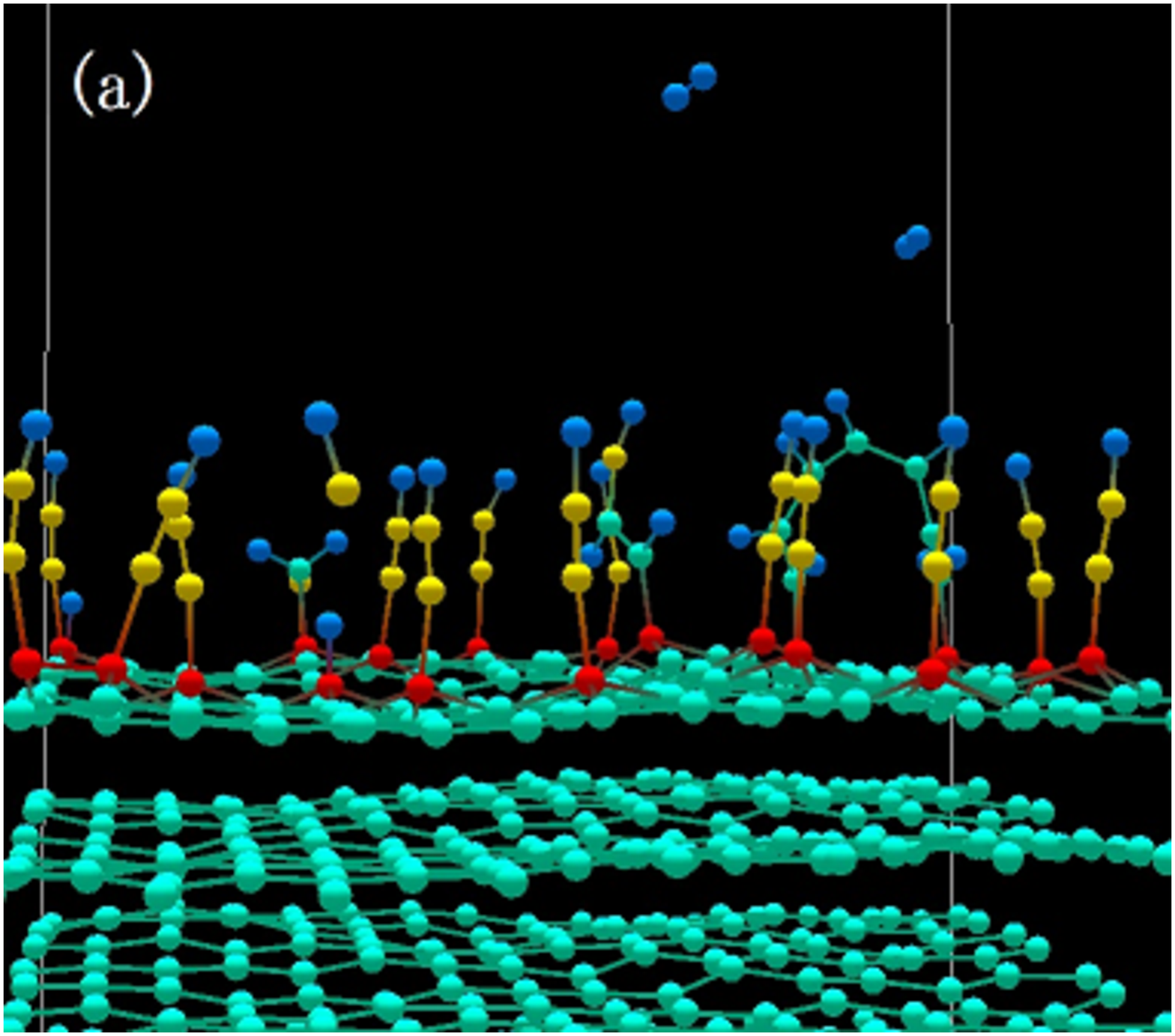}} &
		\resizebox{0.5\linewidth}{!}{\includegraphics{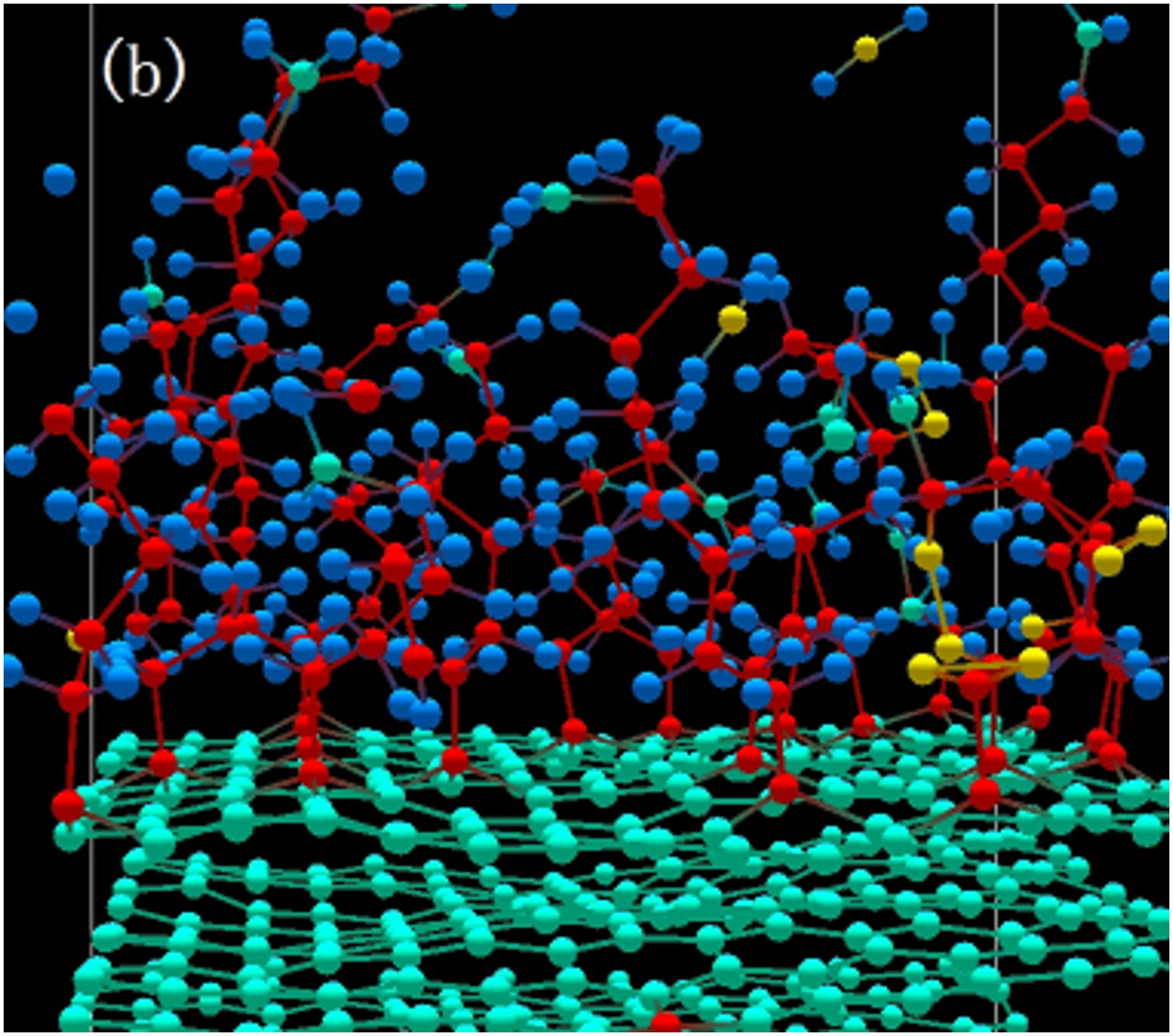}} \\
		\resizebox{0.5\linewidth}{!}{\includegraphics{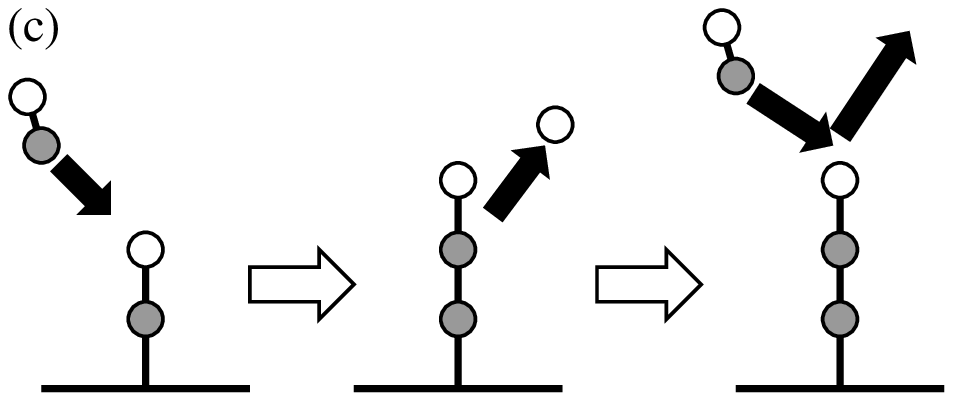}} &
		\resizebox{0.5\linewidth}{!}{\includegraphics{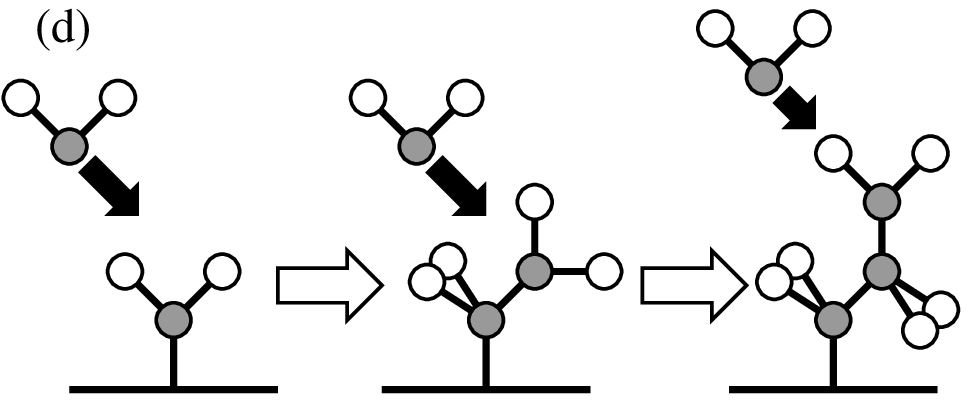}} \\
	
	\end{tabular}
	\caption{~}
	\label{fig:cvd5}
\end{figure}

\begin{list}{}{}

\item[Fig. \ref{fig:method1}.] {(Color online) MD simulation system for PSI.}

\item[Fig. \ref{fig:cvd6}.] {(Color online) Photograph of the MD simulation on the diamond (100) surface. The yellow, green, and red spheres indicate the sp$^1$, ap$^2$, and sp$^3$ carbon atoms, respectively. The blue spheres indicate carbon atoms that have one covalent bond. The purple spheres indicate carbon atoms connected by five covalent bonds or more.   }

\item[Fig. \ref{fig:cvd2kai}.] { {(a) Deposition rates of the carbon atoms on the diamond (100) and (111), and graphite (0001) surfaces as functions of the source H/C ratio and (b) its enlargement. Data points under the dashed line of 0.064 in (b) are depositions less than the number of carbon atoms corresponding to a monolayer. (c) Experimental data cited from Liu et al.\cite{Liu}}}

\item[Fig. \ref{fig:cvd3}.] {The compositions of the carbon deposits for the bonding state of carbon atoms were calculated as functions of the source H/C ratio: (a) a diamond (100) surface, (b) a diamond (111) surface, and (c) a graphite (0001) surface.}

\item[Fig. \ref{fig:cvd4}.] {Deposition rates of carbon atoms in the cases of the hydrocarbon molecules: (a) a diamond (100) surface, (b) a diamond (111) surface, and (c) a graphite (0001) surface.}

\item[Fig. \ref{fig:cvd5}.] {{(Color online) Photographs of MD simulation in the cases of (a) CH and (b) CH$_2$ injections. The blue, yellow, green, and red spheres indicate atoms having one to four covalent bonds, respectively.  Typical deposition processes of (c) CH and (d) CH$_2$. The white and gray spheres indicate hydrogen and carbon atoms, respectively.}}

\end{list}

\clearpage
\thispagestyle{empty}

~
\end{document}